\documentclass[aps,prl,groupedaddress,twocolumn,showpacs,preprintnumbers,superscriptaddress]{revtex4}
\usepackage{graphicx,amsmath,amssymb,bm}

\begin{document}
\title{Spin-polarized tunneling microscopy and the Kondo effect}
\author{Kelly R. Patton}
\affiliation{I. Institut f\"ur Theoretische Physik Universit\"at Hamburg, Hamburg 20355, Germany }
\author{Stefan Kettemann}
\affiliation{I. Institut f\"ur Theoretische Physik Universit\"at Hamburg, Hamburg 20355, Germany }
\author{Andrey Zhuravlev}
\affiliation{I. Institut f\"ur Theoretische Physik Universit\"at Hamburg, Hamburg 20355, Germany }
\affiliation{Institute of Metal Physics, Ekaterinburg 620219, Russia}
\author{Alexander Lichtenstein}
\affiliation{I. Institut f\"ur Theoretische Physik Universit\"at Hamburg, Hamburg 20355, Germany }
\date{\today}
\begin{abstract}
We present a theory for spin-polarized scanning tunneling microscopy (SP-STM) of a Kondo impurity on an unpolarized metallic substrate.  The spin polarization of the SP-STM  breaks the spin symmetry of the Kondo system, similar to an applied magnetic field, leading to a splitting of the Abrikosov-Suhl-Kondo resonance.  The amount of splitting is  controlled by the strength of the coupling between the impurity and the SP-STM tip and also the overall spin polarization of the SP-STM.
\end{abstract}
\pacs{72.15.Qm,72.25.-b}
\maketitle

The Kondo effect has become one of the hallmarks of many-body physics \cite{Hewson}, stimulating  development of both new experimental and theoretical techniques.  Apart from its original manifestation, in the form of anomalous  low-temperature resistance of metals with magnetic impurities, its signature has been observed in a variety of other systems, such as transport in quantum dots \cite{GoldhaberNatrue98,CronenwettScience98} and, more recently, by scanning tunneling microscopy (STM) of  single magnetic adatoms on a metallic surface \cite{LiPRL98,MadhavenScience98,KnorrPRL02}. In these experiments, below the Kondo temperature, an enhanced conductance near the Fermi energy is found due to the formation of a large peak  in the density of states, the so-called  Abrikosov-Suhl-Kondo (ASK) resonance.  The imaging of a Kondo impurity with an STM does not directly resolve the ASK resonance;  a more complex feature is found \cite{LiPRL98,MadhavenScience98,PlihanPRB01,NagaokaPRL02}. This feature is similar to the Fano resonance \cite{FanoPR61}, more commonly found in atomic physics, and can be explained as resulting from an interference of two tunneling paths; one from the STM tip to the substrate and the other from the tip to the impurity-atom then to the substrate.

An STM allows one to study a Kondo system at the atomic level, and with the advent of the spin-polarized STM (SP-STM) \cite{WiesendangerPRL90,HeinzeScience00}, it is now possible to  study spin resolved aspects of the Kondo effect.  Because spin-dynamics is at the heart of the Kondo effect, the use of an SP-STM as a probe seems natural. However, owing to the spin symmetry of a Kondo system, one might expect nothing of interest to be found. Although, the spin symmetry is only preserved if one neglects the effects of the SP-STM tip on the Kondo system. Including the effects of the spin-polarized current on the Kondo impurity, in the language of the Anderson model \cite{AndersonPR61}, adds a spin dependent hybridization term to the Hamiltonian.  We show this spin dependent hybridization mimics an applied magnetic field, splitting the ASK resonance of the impurity. The splitting of the ASK resonance is similar  to that found in quantum dots coupled to ferromagnetic leads/baths \cite{ChoiPRL2004,MartinekPRL2003,PasupathyScience04}.
We find that  with an SP-STM   a Fano like line-shape is observed, but because of the coupling of the impurity to the tip---leading to the splitting of the ASK resonance---the Fano line-shape also exhibits this splitting.

\begin{figure}
\includegraphics[scale=.34]{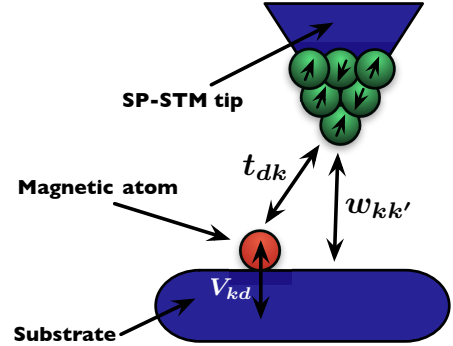}%
\caption{\label{fig1} (Color online) Model of a Kondo system coupled to an SP-STM.}
\end{figure}

{\it Model.}---We model the combined Kondo and SP-STM system as a single impurity on a noninteracting conducting substrate; both are coupled by tunneling to the SP-STM, typically an antiferromagnetic coated standard STM tip, see Fig.~\ref{fig1}. 
 The full Hamiltonian  is  ($\hbar =1$) $H=H_{\rm And}+H_{\rm tip}+H_{\rm tun},$
where
\begin{align}
H_{\rm And}&=\sum_{k,\sigma}\epsilon^{}_{k}c^{\dagger}_{k\sigma}c^{}_{k\sigma}+\sum_{\sigma}E_{d\sigma}d^{\dagger}_{\sigma}d^{}_{\sigma}+Un_{\uparrow}n_{\downarrow}\nonumber \\ &+\sum_{k,\sigma}V^{}_{kd}c^{\dagger}_{k\sigma}d^{}_{\sigma}+{\rm H.c.}
\end{align}
is the single-impurity Anderson model \cite{AndersonPR61},  describing an impurity level $E_{d\sigma}$ with on-site Coulomb term $U$ along with  hybridization $V_{kd}$ with the substrate.  The Hamiltonian for the SP-STM is taken to be
\begin{equation}
H_{\rm tip}=\sum_{k,\sigma}\epsilon'^{}_{k\sigma}a^{\dagger}_{k\sigma}a^{}_{k\sigma},
\end{equation}
where $\epsilon'_{k\sigma}=\epsilon_{k\sigma}+eV$. For the tip we allow for a different spectrum, therefore a different density of states, for each spin.  The charge of the electron is $-e$ and $V$ is the applied voltage.    Because tunneling can occur  from the tip to the impurity, with amplitude $t^{\sigma}_{dk}$, or to the substrate, with amplitude $w^{\sigma}_{kk'}$, both processes have to be included in the tunneling term
\begin{equation}
H_{\rm tun}=\sum_{k,\sigma}t^{\sigma}_{dk}d^{\dagger}_{\sigma}a^{}_{k\sigma}+{\rm H.c.}+\sum_{k,k',\sigma}w^{\sigma}_{kk'}c^{\dagger}_{k\sigma}a^{}_{k'\sigma}+{\rm H.c.}
\end{equation}
We allow for the tunneling barrier to have spin dependance, inasmuch as the tunneling amplitudes  $t^{\sigma}_{dk}$ and $w^{\sigma}_{kk'}$ can be different for respective spins, but we do not allow the barrier to flip spins. For an SP-STM, this is known experimentally \cite{dingEurophyslett,bodepivcom} and introduces a tip-system separation dependance on the spin polarization of the tunneling current. We therefore define the polarization of the tunneling current as $P_{t}=(\Gamma^{\uparrow}_{t}-\Gamma^{\downarrow}_{t})/(\Gamma^{\uparrow}_{t}+\Gamma^{\downarrow}_{t})$, where $\Gamma^{\sigma}_{t}=\pi \sum_{k}|t^{\sigma}_{dk}|^{2}\delta(\omega-\epsilon^{}_{k\sigma})$.   We will also make use of the equilibrium Hamiltonian $H_{0}\equiv H(eV=0)$. 
In principle,  super-exchange/RKKY effects between tip and the impurity atom are present.  We address these interactions in later publications and focus, here,  solely on the effects of a spin-polorized current on the Kondo system. 

{\it Conductance.}---Using the number operator for tip electrons,
$N_{\rm tip}=\sum_{k,\sigma}a^{\dagger}_{k\sigma}a^{}_{k\sigma}$, the tunneling current is
 $I=-e\big<\partial_{t}N_{\rm tip}\big>_{H}$. This expectation value could be evaluated using the standard tunneling Hamiltonian formalism \cite{Mahanbook}; however,  if vertex corrections are neglected, this only accounts for the interaction of the SP-STM and the Kondo system to lowest order.  Here we follow Ref.~[\onlinecite{PlihanPRB01}] and use nonequilibrium  Green's function methods. This allows one to treat the applied voltage as the small parameter of perturbation theory, instead of the SP-STM-system coupling and the voltage, as is normally done. Assuming the density of states of the tip $\rho_{\rm tip}$ and the bare substrate $\rho_{\rm sub}$ are energy independent, the total current can be written as $I=I_{\rm sub}+\delta I$, with
$I_{\rm sub}=2e\pi  \rho_{\rm sub}eV\sum_{\sigma}|w^{\sigma}|^{2} \rho^{\sigma}_{\rm tip}$,
which describes tunneling into a bare substrate, along with an additional term given by 
\begin{align}
&\delta I=2e\pi^{2} |V|^{2} \rho^{2}_{\rm sub}\int d\omega\, \big[n^{\rm F}_{\rm tip}(\omega+eV)-n^{\rm F}_{\rm sub}(\omega)\big]\nonumber \\ &\times\sum_{\sigma}|w^{\sigma}|^{2}\rho^{\sigma}_{\rm tip}\big\{(1-q_{\sigma}^{2})\,{\rm Im}\,[G^{d}_{\sigma}(\omega)] +2q_{\sigma}\, {\rm Re}\, [G^{d}_{\sigma}(\omega)]\big\},
\end{align}
where 
\begin{equation}
\label{fano factor}
q_{\sigma}=\frac{t^{\sigma}+w^{\sigma}V{\rm Re}\,[G^{R}_{0}(\omega)]}{\pi w^{\sigma}V \rho_{\rm sub}(\omega)},
\end{equation}
is a spin dependent Fano parameter with $G^{R}_{0}(\omega)=\sum_{k}(\omega-\epsilon_{k}+i\eta)^{-1}$
and where $G^{d}_{\sigma}(\omega)$ is the Fourier transform of the retarded impurity Green's function $G^{d}_{\sigma}(t)=-i\theta(t)\big<\{d^{}_{\sigma}(t),d^{\dagger}_{\sigma}(0)\}\big>_{H_{0}}$.
It should be noted that the time dependance and expectation value of the impurity's Green's function are with respect to the full {\it equilibrium} Hamiltonian $H_{0}$;
$O(t)=e^{iH_{0}t}Oe^{-iH_{0}t}\hspace{2mm} \text{and}\hspace{2mm} \big< O\big>_{H_{0}}={\rm Tr}\,e^{-\beta H_{0}O}/{\rm Tr}\,e^{-\beta H_{0}}$. 
The total tunneling current contains the direct tunneling into the substrate and the impurity but also has  an additional quantum interference term. It is this additional term that leads to a Fano line-shape in the conductance instead of the ASK resonance. 
The differential conductance ${\sf G}\equiv dI/d(eV)$ is then
\begin{equation}
\label{conductance1}
{\sf G}(\omega)=\sum_{\sigma}G^{\sigma}_{\rm sub}\left[1+Y_{\sigma}(\omega)\right]\end{equation}
\footnote{Because of the interference of tunneling paths, Eq.~(\ref{conductance1}) is {\it not} proportional to a density of states, as is normally the case in tunneling \cite{Mahanbook}.},
where $G^{\sigma}_{\rm sub}=2e\pi  \rho_{\rm sub}|w^{\sigma}|^{2}\rho^{\sigma}_{\rm tip}$ is the direct tunneling conductance between the tip and bare substrate, 
$Y_{\sigma}(\omega)=\Gamma_{V}\left\{(1-q_{\sigma}^{2})\,{\rm Im}\,[G^{d}_{\sigma}(\omega)] +2q_{\sigma}\, {\rm Re}\, [G^{d}_{\sigma}(\omega)]\right\}$,
and $\Gamma_{V}=\Gamma^{\uparrow}_{V}=\Gamma^{\downarrow}_{V}=\pi |V|^{2}\rho_{\rm sub}$ is the level-broadening of the $d$-level due to the coupling  $V$ to the substrate.   
For energies close to the Fermi energy Eq.~(\ref{conductance1}) can be recast into the well known Fano line-shape giving      
\begin{equation}
{\sf G}(\omega)\sim\sum_{\sigma}G^{\sigma}_{\rm sub}\frac{[q_{\sigma}+\varepsilon_{\sigma}(\omega)]^{2}}{1+\varepsilon_{\sigma}^{2}(\omega)},
\end{equation}
with rescaled energy  $\varepsilon_{\sigma}=(\omega-E_{d\sigma}+{\rm Re}\, \Sigma^{d}_{\sigma})/{\rm Im}\, \Sigma^{d}_{\sigma}$ ($ \Sigma^{d}_{\sigma}$ is the self-energy of the impurity).
Here we treat the Fano factor, Eq.~(\ref{fano factor}),  as an energy independent fit parameter; therefore, all of the energy dependance is in the impurity's Green's function.

\begin{figure}
\includegraphics[scale=.36]{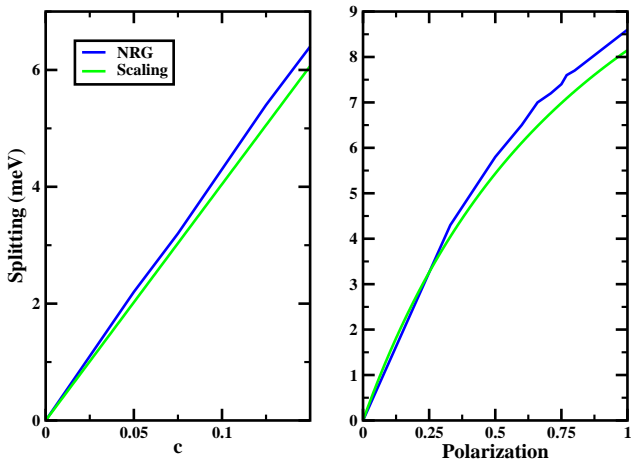}%
\caption{\label{fig2} (Color online)  Shown is  the value of the splitting of the ASK resonance (taken as the difference of the spin up and spin down peaks in the density of states), for (a) a fixed spin polarization of $P=1/3$ spin up and (b)  for a fixed tip-impurity coupling $c=\Gamma^{\uparrow}_{t}/\Gamma_{V}=.10$, as a function of the  polarization of the tunneling current. The remaining  parameters are: $\Gamma_{V}=.2\,{\rm eV}$$, E_{d}=-.9\,{\rm eV}$, $D_{0}=5.5\,{\rm eV},$ and $U=2.9\, {\rm eV}$. }
\end{figure}

{\it Calculation of $G_{\sigma}^{d}$}.---To calculate the impurity's Green's function a variety of methods are available, including equation-of-motion \cite{MartinekPRL03} or numerical renormalization group \cite{ChoiPRL2004,MartinekPRL2003}. Let us first use a simpler and, we believe a more, transparent method. For the asymmetric Anderson model, it is known the ASK resonance splits in the presence of  spin-polarized leads/baths  \cite{ChoiPRL2004,MartinekPRL2003,PasupathyScience04}. (Surprisingly  the  symmetric Anderson model exhibits no such splitting \cite{ChoiPRL2004}.) First we use scaling equations to determine the amount of this splitting \cite{MartinekPRL03,Matsubayashiarxiv07} for our model and then simply represent the ASK resonances as Lorentzians \cite{NagaokaPRL02} centered away from the Fermi energy by the  splitting value. These approximations are  good for energies close to the Fermi energy and weak tip-impurity coupling, where asymmetries of the ASK peak remain small. For example see Fig.~\ref{fig2}.   These asymmetries, such as the widths and heights of the peaks, become more pronounced \cite{ChoiPRL2004,RoschPRB03} and more important for larger splittings than those predicted for our model.   
 
For this calculation (of the impurity's Green's function) we neglect the tunneling into the substrate \footnote{This neglects higher order interference effects and for the symmetric flat band model is negligible.}.  This reduces the full Hamiltonian  to that of a single impurity coupled to two baths/leads; one being the substrate and the other the SP-STM tip. This two-lead-Kondo system can be down folded to a single lead (with spin-dependent hybridization) by the following canonical transformation, $c_{k\sigma}=(|V_{kd}|^{2}+|t^{\sigma}_{dk}|^{2})^{-1/2}(V_{kd}f_{k\sigma}-t^{\sigma}_{dk}h_{k\sigma})$ and $a_{k\sigma}=(|V_{kd}|^{2}+|t^{\sigma}_{dk}|^{2})^{-1/2}(t^{\sigma}_{dk}f_{k\sigma}+V_{kd}h_{k\sigma})$.  This transformation requires the density of states of the leads be equal; therefore, we impose the spin polarization entirely by the tunneling matrix elements $t^{\sigma}_{dk}$.  The splitting of the ASK resonance can be understood as coming from the spin dependent   renormalization of the bare energy levels $E_{d\sigma}$, analogous to an applied magnetic field.  Using poor-man's scaling the renormalized energies levels \cite{AndersonJPhysC70,HaldanePRL78,MartinekPRL03,Matsubayashiarxiv07}, in our model, are  given by 
\begin{align}
&\tilde{E}_{d\uparrow}=E_{d\uparrow}+\frac{\Gamma_{V}}{\pi}\left(1+c\right)\ln\left(\frac{D_{0}}{D_{1}}\right)\nonumber \\ \nonumber {\rm and}\\ 
&\tilde{E}_{d\downarrow}=E_{d\downarrow}+\frac{\Gamma_{V}}{\pi}\left[1+c\left(\frac{1-P_{t}}{1+P_{t}}\right)\right]\ln\left(\frac{D_{0}}{D_{1}}\right),
\end{align}
where we have chosen (arbitrarily) a spin up polarized current, with $P_{t}\in[0,1]$, $c=\Gamma^{\uparrow}_{t}/\Gamma_{V}$, and where $D_{0}$ is the bandwidth cutoff and $D_{1}$ is the  reduced bandwidth. For most adatom systems, $U<D_{0}$, unlike most quantum dot systems where $U\gg D_{0}$. Thus the scaling equations should be cut-off at $D_{1}\approx U$. Therefore 
\begin{equation}
\label{split ed}
\tilde{E}_{d\downarrow}-\tilde{E}_{d\uparrow}\approx\frac{2\Gamma_{V}}{\pi}\,c\left(\frac{P_{t}}{1+P_{t}}\right)\ln\left(\frac{D_{0}}{U}\right).
\end{equation}
Eq.~(\ref{split ed}) shows the  amount of splitting is linear in the coupling strength $c$ but is also a function of the  spin polarization $P_{t}$. 
Fig.~\ref{fig2} shows the range of the splitting.  For comparison, we also plot the splitting as calculated using numerical  renormalization group (NRG).  We have chosen a parameter set to reflect common experimental values \cite{LinPRL06}; although, the Kondo temperature derived from these particular parameters is much larger than typical systems.

Near the Fermi energy, the spin resolved density of states of the impurity can be well approximated by  Lorentzians \cite{NagaokaPRL02} which, here, are centered away from the Fermi energy by the amount of splitting, Eq.~(\ref {split ed}),
\begin{align}
\label{dos}
\rho_{\pm}^{d}(\omega) & =-\frac{1}{\pi}\,{\rm Im}[G^{d}_{\pm}(\omega)]\nonumber\\&\approx\frac{1}{\pi\Gamma_{V}}\left[1+\left(\frac{\pi T}{\sqrt{2}T_{\rm K}}\right)^{2}\right]^{-1}\nonumber \\ &\times\left\{1+\frac{[\omega\pm(\tilde{E}_{d\downarrow}-\tilde{E}_{d\uparrow})]^{2}}{(\pi k_{\rm B}T)^{2}+2(k_{\rm B}T_{\rm K})^{2}}\right\}^{-1}
\end{align}
where $T_{\rm K}$ is the Kondo temperature  and $k_{\rm B}$ is the Boltzmann constant.  
\begin{figure}
\includegraphics[scale=.355]{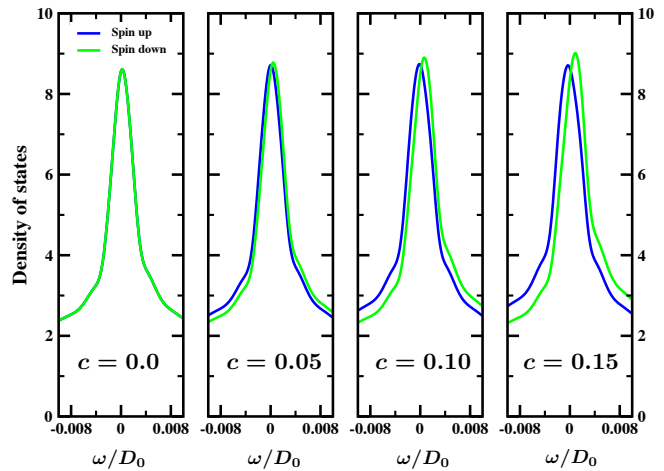}
\caption{\label{fig3} (Color online) The NRG calculated spin-resolved density of states, $\rho^{d}_{\sigma}(\omega)=-1/\pi\,{\rm Im}\,[G^{d}_{\sigma}(\omega)]$, of the Kondo impurity is plotted for different tip-impurity coupling strengths; $c=\Gamma^{\uparrow}_{t}/\Gamma_{V}$. We set $\Gamma^{\downarrow}_{ t}/\Gamma^{\uparrow}_{t}=|t^{\downarrow}|^{2}/|t^{\uparrow}|^{2}=1/2$, which gives a $33\%$  spin up polarized tunneling current. All other parameters are the same as Fig.~\ref{fig2}.}
\end{figure}
\begin{figure}
\includegraphics[scale=.35]{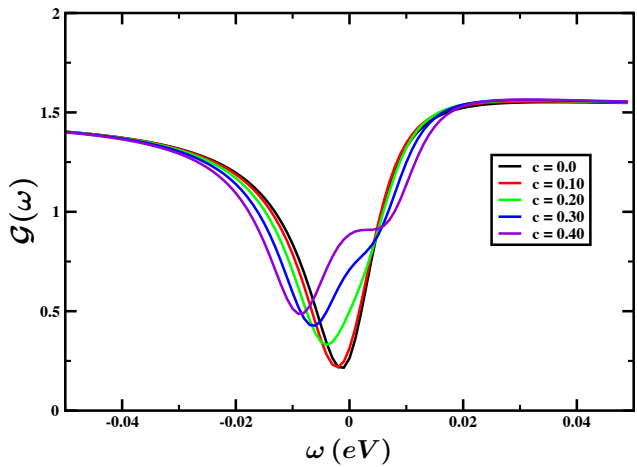}%
\caption{\label{fig4} (Color online) For $T=10\,{\rm K}$, the normalized conductance ${\cal G}(\omega)={\sf G}(\omega)/G^{\uparrow}_{\rm sub}$  is shown for various values of the tip-impurity coupling $c=\Gamma^{\uparrow}_{t}/\Gamma_{V}$ and with $G_{\rm sub}^{\downarrow}/G_{\rm sub}^{\uparrow}=1/2$. For simplicity we set the Fano parameter $q_{\sigma}=.2$  for both spins and impose a $T_{\rm K}=50\, {\rm K}$. Both of which are representative of Co/Cu(111)  \cite{LinPRL06}.  All other parameters are the same as Fig.~\ref{fig2}.}
\end{figure}

{\it Results.}---The Fano parameter, Eq.~(\ref{fano factor}), depends on the microscopic details of the coupling of the impurity to its environment \cite{LinPRL06}. Here we leave it as an unknown fit parameter and choose a fixed value; although, changing the coupling does change the Fano parameter \cite{NeelPRL2007,LinPRL06}. In Fig.~\ref{fig4} we plot the enhancement of the conductance (The real part of the impurity's Green's function can be found by the Hilbert transform of Eq.~(\ref{dos})). As one might expect the splitting of the ASK resonance leads to a splitting of the Fano line-shape.

We have explored the use of an SP-STM to probe the Kondo effect.  By including the effect of the SP-STM on the Kondo system, we find that the spin-polarized current of the SP-STM adds a spin dependent hybridization term to the standard Anderson model.  This hybridization both broadens and renormalizes the bare energy levels, of the impurity---in a spin dependent way---which in turn leads to a splitting of the ASK resonance. This splitting of the ASK resonance is seen as a splitting of the Fano line-shape of the conductance.  The amount of splitting can be controlled by the tip-adsorbate coupling and the overall spin polarization of the SP-STM. Although the splitting is relatively small, to obtain an equivalent splitting with the use of a magnetic field would require field strengths of order $10^{2}$ Tesla. 
 The ability to detect such an effect requires the width of the ASK resonance, which is related to the Kondo temperature,  to be of order of the splitting or smaller.  All things being equal, systems  such as Ti/Ag(100)  with a $T_{\rm K}\approx40\,{\rm K}$ \cite{NagaokaPRL02} or smaller would probably be needed.

Future investigations could include extending the model to include super-exchange/RKKY effects between the tip and the impurity, these become increasing important near the contact regime \footnote{The RKKY   coupling  is of
 order $J_{\rm RKKY} \sim (V^2/U) (w^2/U_{\rm tip})/\epsilon_{\rm F}$. With typical on site interaction  $U_{\rm tip} \le 5\, {\rm eV}$ and for $w < V$ we find  $J_{\rm RKKY} < 1\, {\rm meV}$ for the parameters considered here.}.  Replacing the non-spin-polarized substrate with a ferromagnetic one, introduces further spin dependent coupling.   Experiments such as these could continue to test the effect of spin polarization of the bath(s) on the Kondo effect.  Several theoretical predictions have been made in this area, \cite{ChoiPRL2004,MartinekPRL2003} but few experiments have been performed. \cite{PasupathyScience04} Also it is believed that the tunneling barrier depends on spin for an SP-STM; although, the exact nature of this is unknown and is currently an open question. This dependence  could be experimentally observed by measuring the Fano line-shape splitting as a function of the tip-system separation. 
  
This research was supported by the German Research Council (DFG) under SFB 668 and in part by  Grant No. 4640.2006.2 (Support of Scientific School) from the Russian Basic Research Foundation. S.K. would like to thank Pascal Simon for stimulating discussions. 
\bibliography{/Users/kpatton/Bibliographies/Master}

\end{document}